\documentclass[11pt]{article}

\usepackage[margin=1in]{geometry}
\usepackage{amsfonts}
\usepackage{amssymb}
\usepackage{amsmath}
\usepackage{latexsym}
\usepackage{url}
\usepackage[colorlinks=false,breaklinks]{hyperref}
\usepackage{graphicx, amsthm}
\usepackage{color}

\long\def\rem#1{}
\def\01{\{0,1\}}
\newcommand{\bra}[1]{\langle#1|}
\newcommand{\ket}[1]{|#1\rangle}
\newcommand{\braket}[2]{\langle#1|#2\rangle}
\newcommand{\norm}[1]{\| #1 \|}
\newcommand{\NAND}{\mathop{\overline\wedge}}
\renewcommand{\H}{H}

\def\ketbra #1#2{\ket{#1}\!\bra{#2}}

\def\abs #1{\lvert #1\rvert}
\def\norm #1{\lVert #1\rVert}
\newcommand{\beq}{\begin{equation}}
\newcommand{\eeq}{\end{equation}}

\newcommand{\comment}[1]{\emph{\color{blue}Comment:\color{black} #1}} 
\newlength{\commentslength}
\newcommand{\comments}[1]{
\hspace{-2\parindent}
\addtolength{\commentslength}{-\commentslength}
\addtolength{\commentslength}{\linewidth}
\addtolength{\commentslength}{-\parindent}
\fcolorbox{blue}{white}{\smallskip\begin{minipage}[c]{\commentslength}
\emph{Comments:}\begin{itemize}#1\end{itemize}\end{minipage}}\bigskip
}
\renewcommand{\comment}[1]{}\renewcommand{\comments}[1]{}

\newcommand{\captionfonts}{\small}
\makeatletter
\long\def\@makecaption#1#2{%
  \vskip\abovecaptionskip
  \sbox\@tempboxa{{\captionfonts #1: #2}}%
  \ifdim \wd\@tempboxa >\hsize
    {\captionfonts #1: #2\par}
  \else
    \hbox to\hsize{\hfil\box\@tempboxa\hfil}%
  \fi
  \vskip\belowcaptionskip}
\makeatother

\newtheorem{theorem}{Theorem}
\newtheorem{lemma}[theorem]{Lemma}

\newtheorem{claim}[theorem]{Claim}
\newtheorem{definition}{Definition}
\newtheorem{remark}{Remark}

\newcommand{\thmref}[1]{\hyperref[#1]{{Theorem~\ref*{#1}}}}
\newcommand{\lemref}[1]{\hyperref[#1]{{Lemma~\ref*{#1}}}}
\newcommand{\corref}[1]{\hyperref[#1]{{Corollary~\ref*{#1}}}}
\newcommand{\eqnref}[1]{\hyperref[#1]{{(\ref*{#1})}}}
\newcommand{\defref}[1]{\hyperref[#1]{{Definition~\ref*{#1}}}}
\newcommand{\secref}[1]{\hyperref[#1]{{Section~\ref*{#1}}}}
\newcommand{\figref}[1]{\hyperref[#1]{{Figure~\ref*{#1}}}}
\newcommand{\remref}[1]{\hyperref[#1]{{Remark~\ref*{#1}}}}
\newcommand{\claimref}[1]{\hyperref[#1]{{Claim~\ref*{#1}}}}

\DeclareMathOperator{\Span}{\operatorname{span}}
\DeclareMathOperator{\poly}{\operatorname{poly}}
\DeclareMathOperator{\Reflection}{\operatorname{Reflection}}
\allowdisplaybreaks[1]

\begin{document}

\title{Every NAND formula of size $N$ can be evaluated \\
in time $N^{\frac{1}{2}+o(1)}$ on a quantum computer}
\newcommand{\mystrut}{\rule{1in}{0pt}}
\author{%
\mystrut \and
Andrew M.~Childs%
  \thanks{Institute for Quantum Information, California Institute of Technology.
  Supported by NSF Grant PHY-0456720 and ARO Grant W911NF-05-1-0294.}
\\{amchilds@caltech.edu}
\and
Ben W.~Reichardt$^*$
\\{breic@caltech.edu}
\and \mystrut \and \mystrut \and
Robert \v Spalek%
  \thanks{University of California, Berkeley.
  Supported by NSF Grant CCF-0524837 and ARO Grant DAAD 19-03-1-0082.
  Work conducted in part while visiting Caltech.}
\\{spalek@eecs.berkeley.edu}
\and
Shengyu Zhang$^*$
\\{shengyu@caltech.edu}
\and \mystrut
}
\date{}

\maketitle

\begin{abstract}
For every NAND formula of size $N$, there is a bounded-error $N^{\frac{1}{2}+o(1)}$-time quantum algorithm, based on a coined quantum walk, that evaluates this formula on a black-box input.  Balanced, or ``approximately balanced," NAND formulas can be evaluated in $O(\sqrt{N})$ queries, which is optimal.  It follows that the $(2-o(1))$-th power of the quantum query complexity is a lower bound on the formula size, almost solving in the positive an open problem posed by Laplante, Lee and Szegedy.  
\end{abstract}

\section{Introduction}

Consider a formula $\varphi$ on $N$ inputs $x_1, \ldots, x_N$, using the gate set $S$ either $\{\text{AND, OR, NOT}\}$ or equivalently $\{\text{NAND}\}$.  
That is, the formula $\varphi$ corresponds to a tree where each internal node is a gate from $S$ on its children.
If the same variable is fed into different inputs of $\varphi$, we treat each occurrence separately, so that $N$ counts variables with multiplicity.
The variables $x_i$ are accessed by querying a quantum oracle, which we can take to be the unitary operator $O_x: \ket{b, i} \mapsto (-1)^{b x_i} \ket{b, i}$, for $b \in \{0,1\}$ and $i \in \{1, \ldots, N\}$ the control qubit and query index, respectively.
We will show:

\begin{theorem} \label{t:introduction}
After efficient preprocessing (i.e., preprocessing taking $\poly(N)$ steps) of the formula $\varphi$ independent of $x$, $\varphi(x)$ can be evaluated with error $< 1/3$ using $N^{\frac{1}{2}+o(1)}$ time and queries to $O_x$.  
\end{theorem}

Our algorithm is inspired by the recent $N^{\frac12+o(1)}$-time algorithm of Farhi, Goldstone and Gutmann~\cite{fgg:and-or} for the case in which $S = \{\text{NAND}\}$, each NAND gate in $\varphi$ has exactly two inputs and $\varphi$ is {\em balanced}---i.e., $N = 2^n$ and $\varphi$ has depth $n$.  
Our algorithm requires no preprocessing in this special case of a balanced binary NAND tree.  For a balanced, or even an ``approximately balanced" NAND tree, our algorithm requires only $O(\sqrt{N})$ queries (\thmref{t:generaltheorem}), which is optimal.
The correctness of our algorithm will turn on spectral analysis of a Hamiltonian similar to that simulated by Farhi et al., except in general weighted according to the formula's imbalances.

Our algorithm (almost) solves in the positive the open problem posed by Laplante, Lee and Szegedy \cite{lls:formulas}, whether the quantum query complexity squared is a lower bound on the formula size.  \thmref{t:introduction} implies that the formula size of a function $f$ is at least $Q_2(f)^{2-o(1)}$.  
Note too that evaluating an AND-OR tree is the decision version of evaluating a MIN-MAX tree.

\subsection*{Idea of the algorithm}

\begin{figure}
\centering
\includegraphics[scale=.38]{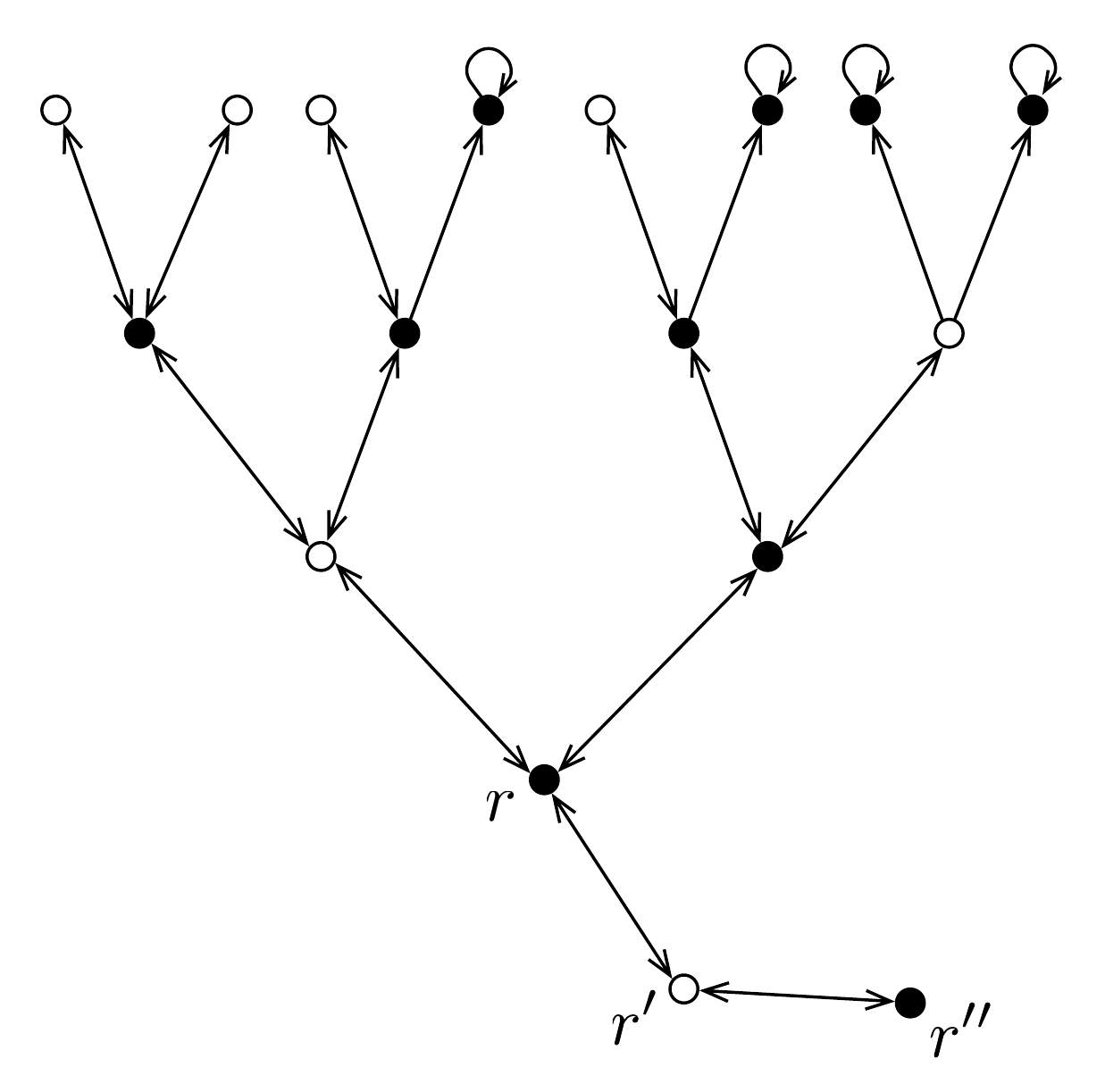}
\caption{The balanced binary NAND formula
$\varphi = \bigl((x_1 \NAND x_2 ) \NAND (x_3 \NAND x_4) \bigr) \NAND \bigl( (x_5 \NAND x_6) \NAND (x_7 \NAND x_8)\bigr)$---where $a \NAND b= \overline {(a \wedge b)}$---is represented by the balanced subtree rooted at $r$.  The input is fed in at the leaves (top row), and internal vertices are evaluated as NAND gates on their children.  Here a vertex $v$ is filled or not according to whether its evaluation $\NAND(v)$ is $1$ or $0$, respectively; in this example, the input is $x=00010111$ and overall, $\varphi(x) = \NAND(r) = 1$.
Below $r$, we have additionally attached two vertices,
$r'$ and $r''$.
Modify the classical uniform random walk on this tree by marking leaf vertices evaluating to $1$ as probability sinks, and adding a certain bias to the coin flip at $r'$.  Then the corresponding quantum walk can be used to evaluate the formula in $O(\sqrt{N})$ time.
} \label{f:shorttailbalancedgraph}
\end{figure}

Associated to the NAND formula $\varphi$ is a tree $T = T(\varphi)$.
\figref{f:shorttailbalancedgraph} gives an example of a balanced binary tree; the formal construction in the general case is in \secref{sec:hamiltonian}.
Consider the classical uniform random walk on this tree; at vertex $v$ with degree $d(v)$, flip an unbiased $d(v)$-sided ``coin" to decide which neighbor to step to next.  (The coin at $r'$ will have a certain bias, as will all coins in the case of an unbalanced tree.)  Incorporate the input into the walk by making all vertices evaluating to $1$ into probability sinks.   

The corresponding quantum walk $U$ works on a vertex register
and a coin register.  Instead of randomizing the coin register
between steps, a Grover diffusion operator is applied to the coin.
This quantum walk $U$, started at $r''$, can be used with
phase estimation to evaluate a general NAND formula in only
$N^{\frac12+o(1)}$ time, or $O(\sqrt{N})$ queries in the balanced or
approximately balanced case.
The general algorithm is described in \secref{sec:algorithm}, while \figref{f:algorithm} presents the full algorithm for the balanced case. 

The correctness proof uses Szegedy's correspondence between classical random walks and discrete-time quantum walks (\thmref{t:szegedization}).  Note that Szegedy's formulation can also be viewed as establishing a correspondence between discrete- and continuous-time quantum walks (\remref{t:continuousdiscretequantumwalkscorrespond}).  In particular, the spectrum and eigenvectors of the coined walk unitary can be related to those of the Hamiltonian for a continuous-time quantum walk on a closely related tree.

\subsection*{History of the problem}

Grover showed in 1996 \cite{grover:search, grover:search-time} how to search a general unstructured database of size $N$, represented by a black-box oracle function, in $O(\sqrt N)$ oracle queries and $O(\sqrt{N} \log \log N)$ time on a quantum computer.  His search algorithm can be used to compute the logical OR of $N$ bits in the same time; simply search for a $1$ in the input string.  Since Grover search can be used as a subroutine, even within another instance of Grover search, one can speed up the computation of more general logical formulas.  For example, a two-level AND-OR tree (with one AND gate of fan-in $\sqrt N$ and $\sqrt N$ OR gates of the same fan-in as its children) can be computed in
$O(\sqrt N \log N)$ queries.  Here the log factor comes from amplifying the success probability of the inner quantum search to be polynomially close to one, so that the total error is at most constant.  By iterating the same argument, regular AND-OR trees of depth $d$ can be evaluated with constant error in time $\sqrt N \cdot O(\log N)^{d-1}$~\cite{BuhrmanCleveWigderson98}.

H\o yer, Mosca and de Wolf \cite{hmw:berror-search} showed that Grover search can be applied even if the input variables are noisy, so the log-factor is not necessary.
Consequently, a depth-$d$ AND-OR tree can be computed in $O(\sqrt N \cdot c^d)$ queries, where $c$ is a constant that comes from their algorithm.  It follows that constant-depth AND-OR trees can be computed in $O(\sqrt N)$ queries.
Unfortunately, their algorithm is too slow for the balanced binary AND-OR tree of depth $\log_2 N$ (although it does give some speedup for sufficiently large constant fan-ins).

Classically, using a technique called alpha-beta pruning, one can compute the value of a balanced binary AND-OR tree with zero error in expected time $O(N^{\log_2[(1 + \sqrt{33})/4]}) = O(N^{0.754})$~\cite{snir:dec, sw:and-or}, and this is optimal even for bounded-error algorithms~\cite{santha:and-or}.    

For a long time, no quantum algorithm was known that performed better than the classical zero-error algorithm, despite the fact that the best known lower bound, from the adversary method, is only $\Omega(\sqrt N)$~\cite{bs:q-read-once}.  Very recently, Farhi, Goldstone and Gutmann~\cite{fgg:and-or} presented a ground-breaking quantum algorithm for the balanced case, based on continuous-time quantum walks.  Their algorithm runs in time $O(\sqrt N)$ in an unconventional, continuous-time query model.  
Childs, Cleve, Jordan and Yeung~\cite{ccjy:and-or} shortly after pointed out that this algorithm can be discretized into the conventional oracle query model with a small slowdown, to run in time $N^{\frac{1}{2} + o(1)}$.

After a previous version of our paper \cite{crsz:andor} was distributed on the preprint arXiv, Ambainis~\cite{ambainis07nand} gave an $O(\sqrt{N})$-query algorithm for evaluating balanced binary AND-OR trees.  This is optimal in the number of oracle queries, and may also be efficiently implementable~\cite{ambainis07personal}.  

\subsection*{Organization}

\secref{sec:hamiltonian} defines and puts weights on the tree $T(\varphi)$.  The weights of edges to the leaves depend on the input $x$.  \secref{sec:zeroenergy} shows that when $\varphi(x) = 0$, there exists a zero-energy eigenstate (i.e., eigenvector with eigenvalue zero) of the weighted adjacency matrix of $T(\varphi)$ and a short tail, having substantial overlap with a known initial state.  Conversely, if $\varphi(x) = 1$, then any zero-energy eigenstates can be neglected, as they have no overlap on the initial state.  In the case $\varphi(x) = 1$, \secref{sec:spectralgapanalysis} shows that eigenvectors with small nonzero eigenvalues can also be neglected.  This is argued by connecting the NAND formula's evaluation to the ratio of amplitudes from a child to its parent.  
We then construct an algorithm using the observations of Sections~\ref{sec:zeroenergy} and~\ref{sec:spectralgapanalysis}.  \secref{sec:szegedization} reviews Szegedy's theorem relating the spectrum and eigenvectors of a coined quantum walk to those of the Hamiltonian for a continuous-time quantum walk.  We apply this relationship in \secref{sec:algorithm} to show that phase estimation on a certain coined quantum walk can be used to evaluate $\varphi$.

\begin{figure}
\begin{center}
\fbox{
\begin{minipage}[l]{6in}
\begin{enumerate}
\item
{\bf Initialization.} Let $T = 320 \lfloor \sqrt N \rfloor$.
Prepare three quantum registers in the state
\[
\Big(\frac{1}{\sqrt T}\sum_{t = 0}^{T-1} (-i)^t \ket t\Big) \otimes \ket{r''} \ket{\mathrm{left}} \enspace.
\]
The first register is a counter for quantum phase estimation, the second register holds a vertex index, and the third register is a qutrit ``coin" holding `down', `left' or `right' in this order.
\item
{\bf Quantum walk.} If the first register is $\ket t$, perform $t$
steps of the following discrete-time coined quantum walk $U$.
Denote the last two registers by $\ket v \ket c$.
\begin{itemize}
\item
{\bf Diffusion step.}
\begin{enumerate}
\item
If $v$ is a leaf, apply a phase flip $(-1)^{x_v}$ using one controlled
call to the input oracle.
\item
If $v$ is an internal degree-three vertex, apply the following diffusion
operator on coin $\ket c$:
\[
\Reflection_{\ket u} = 2 \ket u \bra u - I =
\begin{pmatrix}
-1/3 & 2/3 & 2/3\\
 2/3 &-1/3 & 2/3\\
 2/3 & 2/3 &-1/3
\end{pmatrix}
\enspace ,
\]
where $\ket u = \frac 1 {\sqrt 3} (\ket{\mathrm{down}} +
\ket{\mathrm{left}} + \ket{\mathrm{right}}) = \frac 1 {\sqrt 3} (1,
1, 1)^T$.
\item If $v = r'$, apply the following diffusion operator on $\ket c$:
\[
\Reflection_{\ket{u'}} = 2 \ket{u'} \bra{u'} - I =
\begin{pmatrix}
\frac 2 {\sqrt N} - 1 & 2 \sqrt{\frac 1 {\sqrt N} - \frac 1 N} & 0 \\
2 \sqrt{\frac 1 {\sqrt N} - \frac 1 N} & 1 - \frac 2 {\sqrt N} & 0 \\
0 & 0 & -1 \\
\end{pmatrix}
\enspace ,
\]
where $\ket{u'} = \frac 1 {\sqrt[4]N} \ket{\mathrm{down}} + \sqrt{1 -
\frac 1 {\sqrt N}} \ket{\mathrm{left}}$.
\item
If $v = r''$, do nothing.
\end{enumerate}
\item
{\bf Walk step.}
\begin{enumerate}
\item
If $c =$ `down', then walk down to the parent of $v$ and set $c$ to
either `left' or `right' depending on which child $v$ is.
\item
If $c \in$ \{`left', `right'\}, then walk up to the corresponding
child of $v$ and set $c$ to `down'.
\end{enumerate}
Note that the walk step operator is a permutation that simply flips the direction of each oriented edge.  
\end{itemize}
\item
{\bf Quantum phase estimation.} Apply the inverse quantum Fourier transform
(modulo $T$) on the first register and measure it in the computational basis.
Return $0$ if and only if the outcome is $0$ or $\frac T 2$.
\end{enumerate}
\end{minipage} }
\end{center}
\caption{An optimal quantum algorithm to evaluate the balanced binary
NAND formula using $O(\sqrt N)$ queries.  The algorithm runs quantum
phase estimation on top of the quantum walk of
\figref{f:shorttailbalancedgraph}.} \label{f:algorithm}
\end{figure}

\section{Weighted NAND formula tree} \label{sec:hamiltonian}

Consider a NAND formula of size $N$---i.e., on $N$ variables, counting multiplicity.  (The NAND gate on inputs $y_1, \ldots, y_k \in \{0,1\}$ evaluates to $1 - \prod_{i=1}^k y_i$.  In particular, the NAND gate on a single input is simply the NOT gate.)

Represent the
formula $\varphi$ by a rooted tree $T = T(\varphi)$, in which the leaves correspond to variables, and other vertices correspond to NAND gates on their children.  (Because $\varphi$ is a formula, not a circuit, each gate has fan-out one, so there are no loops in the associated graph.)  
Attach below the root $r$ a ``tail" of two vertices $r'$ and $r''$ as in \figref{f:shorttailbalancedgraph}.

\begin{definition}
For a vertex $v$, let $s_v$ be the number of 
inputs to the subformula under $v$, counting multiplicity (in particular $s_r = s_{r'} = s_{r''} = N$).
Let $\NAND(v)$ denote the value of the subformula at $v$
($\NAND(r) = \NAND(r'') = \varphi(x)$).
\end{definition}

\def\pathsum #1{{\sigma_+({#1})}}
\def\pathinvsum #1{{\sigma_-(#1)}}
\def\path {\xi}

\rem{
\begin{figure}
\centering
\includegraphics[scale=.38]{images/oracleexample}
$\qquad$
\includegraphics[scale=.38]{images/oracleexample2}
\caption{Left: As an example of the correspondence $(\varphi, x) \leftrightarrow T$, the above tree corresponds to the NAND formula $\varphi = \overline{\left(\overline{(x_1 \wedge x_2)} \wedge \overline{(x_3 \wedge x_4)} \wedge x_5 \wedge \overline{(x_6 \wedge \overline{(x_7 \wedge x_8)} \wedge x_9)}\right)} \wedge \overline{(x_{10} \wedge x_{11} \wedge x_{12})}$ evaluated on the input $x_1 \ldots x_{12} = 000110010110$.  The input is reflected by the presence or absence of the bold or dotted edges; for example $e_1$ is not present because $x_7=0$, while the presence of $e_2$ indicates $x_8=1$.  Leaf vertices evaluate to $0$, while each internal vertex evaluates the NAND of its children (or a NOT gate on a single child).  A vertex is filled or not according to whether it evaluates to $1$ or $0$, respectively; overall, $\varphi(x) = \NAND(r) = 0$.  Right: Flipping $x_5$ by removing an edge, the tree now evaluates to $1$.} \label{f:nandexample}
\end{figure}
}

\def \tensor {\otimes}

\begin{definition}
\label{def:h}
Let $\H$ be a symmetric, weighted adjacency matrix of the graph consisting of $T$ and the attached tail:
\begin{equation} \label{eqn:h}
\H \ket v = h_{pv} \ket p + \sum_c h_{vc} \ket c \enspace ,
\end{equation}
where $p$ is the parent of $v$ and the sum is over $v$'s children.  (If $v$ has no parent or no children, the respective terms are zero.)
The edge weights depend on the structure of the tree, and are given by
\begin{equation} \label{eqn:weights}
h_{pv} = \Big(\frac {s_v} {s_p}\Big)^{1/4} \enspace ,
\end{equation}
with two exceptions:
\begin{enumerate}
\item \label{e:hleaf}
If a leaf $v$ evaluates to $\NAND(v) = 1$, set $h_{pv} = 0$, i.e., effectively remove the edge $(p,v)$ by setting its weight to zero.
\item
Set $h_{r''r'} = 1 / (\sqrt{\pathinvsum{r}} N^{1/4})$ (see \defref{t:pathsum}).
\end{enumerate}
\end{definition}

\begin{definition}
\label{t:pathsum}
\label{t:approxbalancedef}
To track error terms through the analysis, it will be helpful to define
\begin{equation} \label{e:pathsum}
\begin{split}
\pathinvsum{v} &= \max_\path
\sum_{\substack{w \in \path:\\ \NAND(w) = 0}} \frac1{\sqrt{s_w}} \\
\pathsum{v} &= \max_\path
\sum_{\substack{w \in \path:\\\NAND(w) = 1}} s_w
\end{split}
\end{equation}
with the maximum in each case taken over all paths $\path$ from $v$ up to a leaf.  
Letting $d_r$ be the depth of $r$, 
clearly $\pathinvsum{v} \leq \pathinvsum{r} \leq d_r$ and $\pathsum{v} \leq \pathsum{r} \leq N d_r$.\footnote{In fact, $\pathinvsum{r} = O(\sqrt{d_r})$, because $s_w$ must increase by at least one every two levels down (two NOT gates in a row would be redundant).  Slightly stronger bounds can be given for trees preprocessed according to the rebalancing procedure of \lemref{t:rebalance}, but $\poly(d_r)$ and $\poly(\log N)$ factors here won't significantly change the running time.}  We call formula $\varphi$ ``approximately balanced'' if $\pathinvsum{r} = O(1)$, $\pathsum{r} = O(N)$ and $\norm{\H} = \norm{\H}_2 = O(1)$.
\end{definition}

Our algorithm will depend on the following properties of $\H$, which we will prove in the following two sections:

\begin{theorem} \label{t:eigensystem}
If $\varphi(x) = 0$, then there exists $\ket a$ a zero-energy eigenvector of $\H$, with $\norm{\ket a} = 1$ and overlap $\braket{r''}{a} \geq 1/\sqrt{2}$.
If $\varphi(x) = 1$, then every eigenvector with support on ${r'}$ or ${r''}$ has corresponding eigenvalue at least $\frac{1}{9 \pathinvsum{r} \sqrt{\pathsum{r}}}$ in absolute value.
\end{theorem}

\begin{remark} \label{t:alternateweights}
For a leaf $v$ evaluating to $0$, it is sufficient that $h_{pv}$ satisfy $h_{pv} \geq 1 / s_p^{1/4}$.
One can also verify \thmref{t:eigensystem} for weights $h_{pv}$ defined by, for an arbitrary fixed $\beta \geq 0$, $h_{pv} = {s_v}^\beta / {s_p}^{\frac12-\beta}$, 
$h_{r''r'} = 1 / (\sqrt{\pathinvsum{r}} N^{\frac12-\beta})$ 
and $\pathinvsum{v} = \max_\path \sum_{\substack{w \in \path:\\ \NAND(w) = 0}} \frac1{{s_w}^{2\beta}}$.  We fix $\beta = 1/4$ to simplify notation.
\end{remark}

\section{Zero-energy eigenstates of $\H$} \label{sec:zeroenergy}

\begin{figure}
\centering
\includegraphics[scale=.4]{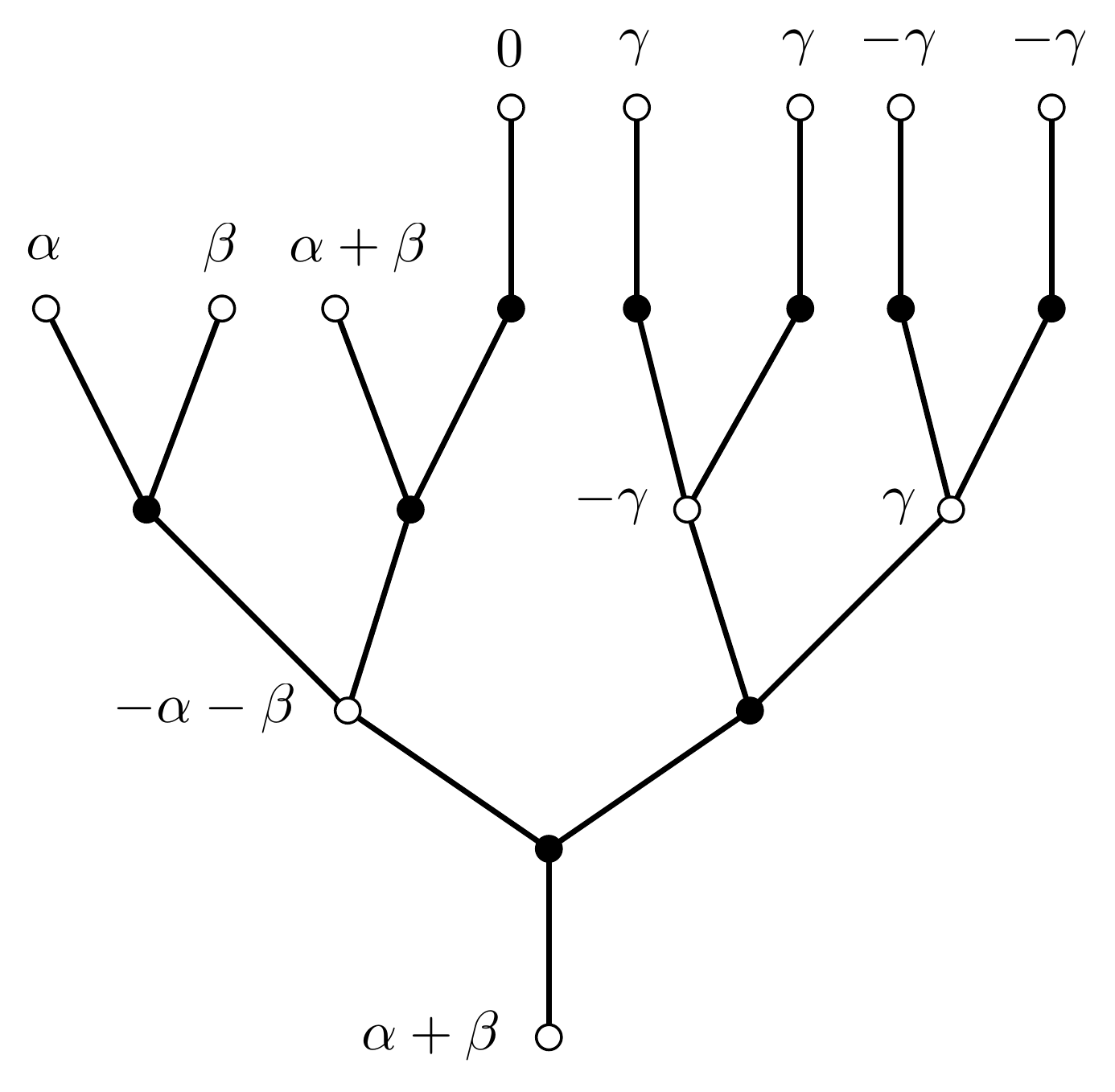}
\caption{An example NAND tree to illustrate Lemmas~\ref{t:energy0forall} and~\ref{t:energy0exists}.  As in
\figref{f:shorttailbalancedgraph},
a vertex is filled or not according to whether it evaluates to $1$ or $0$, respectively.
The amplitudes $\braket{v}{a}$ of a zero-energy eigenstate $\ket{a}$ for the adjacency matrix $\H$ are also labeled, with $\alpha$, $\beta$, $\gamma$ free variables, assuming $h_{pv} = 1$ for every edge $(p,v)$.  The amplitudes of the neighbors of any vertex sum to zero.  The existence of such an $\ket{a}$ is promised by \lemref{t:energy0exists}.  As required by \lemref{t:energy0forall}, $\braket{v}{a} = 0$ if $\NAND(v)=1$, so vertices evaluating to $1$ are not labeled.} \label{f:nand0energyexample}
\end{figure}

Recall that in a NAND tree $T$, internal vertices are interpreted as NAND gates on their children.  
As \defref{def:h} puts zero weight on the parental edge of a leaf evaluating to one, such leaves can be regarded as disconnected.  All leaves connected to the root component can be interpreted as zeros. 
\begin{definition}
By $T_v$, we mean the subtree of $T$ consisting of $v$ and all its descendants.  The restriction to $T_v$ of a vector $\ket{a}$ on $T$ will be denoted $\ket{a_{T_v}}$.  That is, for a subset $S$ of the vertices, define the projector $\Pi_S = \sum_{s \in S} \ketbra{s}{s}$; then $\ket{a_{T_v}} = \Pi_{T_v} \ket{a}$.  We will also write $a_v$ for $\braket{v}{a}$.  Finally, let $\H_{S} = \Pi_{S} \H$.
\end{definition}

\begin{lemma}  \label{t:energy0forall}
For an internal vertex $p$ in NAND tree $T$, if $\NAND(p) = 1$ and $\H_{T_p} \ket{a} = 0$, then $a_p = 0$.  
\end{lemma}

\begin{proof}
Since $\NAND(p) = 1$, there exists a child $v$ of $p$ having $\NAND(v) = 0$.  If $v$ is a leaf, then $0 = \bra{v} \H \ket{a} = h_{pv} a_p$, as asserted.  Otherwise, all children $c$ of $v$ must have $\NAND(c) = 1$, implying by induction that $a_c = 0$.  Then 
$$
0 = \bra{v} \H \ket{a} = h_{pv} a_p + \sum_c h_{vc} a_c = h_{pv} a_p \enspace ,
$$
as asserted.  
\end{proof}

\lemref{t:energy0forall} constrains the existence of zero-energy eigenstates supported on the root $r$ when the NAND formula evaluates to $1$.  
However, there may be zero-energy eigenstates that are not supported on the root (for example, consider the right subtree in \figref{f:nand0energyexample}).

\begin{lemma} \label{t:energy0exists}
Consider a vertex $p$ in NAND tree $T$.
If $\NAND(p) = 0$, then there exists an $\ket{a} = \ket{a_{T_p}}$ with $\H_{T_p} \ket{a} = 0$, $\norm{\ket{a}} = 1$, and
$a_p \geq 1/(\sqrt{\pathinvsum{p}} s_p^{1/4})$.
\end{lemma}

\begin{proof}
Since $\NAND(p) = 0$, for all children $v$ of $p$, $\NAND(v) = 1$.  So each $v$ has a child $c^v$ satisfying $\NAND(c^v) = 0$.  

Construct $\ket{a}$ as follows.  
Set $a_v = 0$ for all children $v$.  
Set $\ket{a_{T_c}} = 0$ for all grandchildren $c \notin \{c^v\}$.  
By induction, for each $v$ construct $\ket{\tilde{a}_{T_{c^v}}}$ satisfying $\norm{\ket{\tilde{a}_{T_{c^v}}}} = 1$, $\H_{T_{c^v}} \ket{\tilde{a}_{T_{c^v}}} = 0$ and $\tilde{a}_{c^v} \geq 1/(\sqrt{\pathinvsum{c^v}} s_{c^v}^{1/4})$.

For each $v$, in order to satisfy $\bra{v} \H \ket{a} = 0$, we need $h_{pv} a_p = - h_{vc^v} a_{c^v}$.  To satisfy all these equations, we {\em rescale} the vectors $\ket{\tilde{a}_{T_v}}$.  Let $a_p = 1$, and let $\ket{a_{T_{c^v}}} = - \tfrac{h_{pv}}{h_{v c^v}}\tfrac{1}{\tilde{a}_{c^v}} \ket{\tilde{a}_{T_{c^v}}}$.
It only remains to verify that $\norm{\ket{a}}^2 \leq \sqrt{s_p} \pathinvsum{p}$, so that renormalizing, $a_p / \norm{\ket{a}} = 1/\norm{\ket{a}}$ is still large.  Indeed,\footnote{We use a note beside a line to help indicate the derivation of the next line.}
\begin{align*}
\norm{\ket{a}}^2 
&= a_p^2 + \sum_v \frac{h_{pv}^2}{h_{v c^v}^2 (\tilde{a}_{c^v})^2} \norm{\ket{a_{T_{c^v}}}}^2 && \tilde{a}_{c^v} \geq 1/(\sqrt{\pathinvsum{c^v}} s_{c^v}^{1/4})
\\
&\leq 1 + \sum_v \frac{h_{pv}^2}{h_{vc^v}^2} \sqrt{s_{c^v}} \pathinvsum{c^v} && h_{pv} = \Big(\frac{s_v}{s_p}\Big)^{1/4},\; h_{v c^v} \ge \Big(\frac{s_{c^v}}{s_v}\Big)^{1/4} \\
&\leq 1+ \sum_v \frac{s_v}{\sqrt{s_p}} \pathinvsum{c^v} && \sum_v s_v = s_p\\
&\leq \sqrt{s_p} \Big(\frac1{\sqrt{s_p}} + \max_v \pathinvsum{c^v}\Big) \\
&\leq \sqrt{s_p} \pathinvsum{p} \enspace . &&\qedhere
\end{align*}
\end{proof}

\noindent (The key step in the above proof, which motivates the choice of weights $h_{pv}$, is $\sum_v s_v = s_p$.)

\lemref{t:energy0exists} is a strong converse of \lemref{t:energy0forall}, as it does not merely assert that $a_p$ can be set nonzero; it also puts a quantitative lower bound on the achievable magnitude.  \lemref{t:energy0exists} lets us say that there {\em exists} an energy-zero eigenstate with large overlap with the root $r$ when $\NAND(r) = 0$.  

Now in the case $\varphi(x) = \NAND(r) = 0$, let us extend $\ket{a_{T_r}}$ from \lemref{t:energy0exists} into a zero-energy eigenvector $\ket{a}$ over the whole graph, to see that the overlap $\abs{\braket{r''}{a}} / \norm{\ket a}$ is large.  In order to satisfy $\H \ket{a} = 0$, we must have $a_{r'} = 0$ and $-a_{r''} = \frac{h_{r'r}}{h_{r''r'}} a_r
= \sqrt{\pathinvsum{r}} N^{1/4} a_r \geq 1$.  Therefore, we lower-bound
\[
\frac{\abs{\braket{r''}{a}}}{\norm{\ket a}}
\geq \frac 1 {\sqrt{1 + \norm{\ket{a_{T_r}}}^2}}
= \frac1{\sqrt 2} \enspace .
\]

\section{Spectral gap of $H$ in the case \texorpdfstring{$\varphi(x) = 1$}{phi(x)=1}} \label{sec:spectralgapanalysis}

To complete the analysis in the $\varphi(x) = 1$ case, we investigate $\H$'s eigenvectors corresponding to energies $E$ close to zero.  In this section, we will show

\begin{theorem} \label{t:spectralgap}
If $\varphi(x) = \NAND(r) = 1$, then $\H$ has no eigenvector with energy $\abs{E} \in \left(0,\frac{1}{9 \pathinvsum{r} \sqrt{\pathsum{r}}}\right]$ and support on $r'$ or $r''$.
\end{theorem}

As $T$ is a bipartite graph, $\H$'s eigenvalues are symmetric around zero.  Let
$$
\ket{E} = \sum_v \alpha_v \ket v
$$
be an eigenvector of $\H$ with eigenvalue $E > 0$.  

From Eq.~\eqnref{eqn:h} we obtain
\begin{equation} \label{e:hh}
\bra v \H \ket E = E \alpha_v = h_{pv} \alpha_p + \sum_c h_{vc} \alpha_c \enspace .
\end{equation}
The analysis depends on the fact that $\alpha_v / \alpha_p$ is either large or small in magnitude depending on whether $\NAND(v) = 0$ or 1.

\newcommand{\z}[1]{\sqrt{s_{#1}}}
\begin{lemma} \label{t:ybounds}
Let $0 < E \le \frac 1 {9 \pathinvsum{r} \sqrt{\pathsum{r}}}$.
For vertices $v \neq r''$ in $T$, define $y_{0v}$ and $y_{1v}$ by
\begin{equation} \label{eqn:y1v}
\begin{split}
y_{0v} &= \frac {1 + \sum_c h_{vc} y_{1c}} {h_{pv}} \\
y_{1v} &= \frac {h_{pv} \z{v}} {\gamma_v} \enspace ,
\end{split}
\end{equation}
where $p$ is the parent of $v$, the sum is over children $c$ of $v$, and $\gamma_v$ is defined by
\begin{align}
\gamma_v &= \Gamma - \Gamma^2 \pathinvsum{v} - E^2 (1+\Gamma') \pathsum{v} \label{e:gammav}\\
\Gamma &= \frac 1 {2 \pathinvsum{r}} \nonumber\\
\Gamma' &= 8 \pathinvsum{r} \enspace . \nonumber
\end{align}
(It holds that $1/\Gamma' \le \gamma_v \le \Gamma$.)
Then for every vertex $v \neq r''$ in $T$, either $\alpha_v = \alpha_p = 0$, or
\begin{equation}
\begin{array}{l r @{\ } l}
\NAND(v) = 0 \quad \Rightarrow & 0 <  \alpha_p/\alpha_v &\leq y_{0v} E \\
\NAND(v) = 1 \quad \Rightarrow & 0 > \alpha_v/\alpha_p &\geq -y_{1v} E \enspace .\\
\end{array}
\end{equation}
\end{lemma}

\begin{proof}
The inequality $\gamma_v \le \Gamma$ is trivial.  To show $\gamma_v \ge 1/ \Gamma'$, use $1 \le \pathinvsum v \le \pathinvsum r$, $\pathsum v \le \pathsum r$, and the assumed upper bound on $E$.

The main proof goes by induction.  
Base case: for every leaf $v$, $\NAND(v)=0$ and by Eq.~\eqnref{e:hh}, $E \alpha_v = h_{pv} \alpha_p$.  Thus either $\alpha_v = \alpha_p = 0$,
or $\alpha_p / \alpha_ v =\frac{E}{h_{pv}} = y_{0v} E$.
Induction step:
\begin{itemize}
\item If $\NAND(v) = 0$, then all children $c$ evaluate to $\NAND(c) = 1$.  
First assume $\alpha_v \neq 0$.  Dividing both sides of Eq.~\eqnref{e:hh} by $\alpha_v$, using the induction hypothesis, and rearranging terms gives
\begin{align*}
\frac{\alpha_p}{\alpha_v} 
&= \frac{1}{h_{pv}} \Bigl(E - \sum_c h_{vc} \frac{\alpha_c}{\alpha_v}\Bigr) \\
&\le \frac{1}{h_{pv}} \Bigl(1 + \sum_c h_{vc} y_{1c} \Bigr) E
= y_{0v} E \enspace .
\end{align*}
The induction hypothesis also gives that $\alpha_p / \alpha_v \geq E / h_{pv} > 0$.
If $\alpha_v = 0$, then the induction hypothesis gives that all $\alpha_c$ are zero, so also $\alpha_p = 0$ by Eq.~\eqnref{e:hh}.

\item If $\NAND(v) = 1$, then there is at least one child $c$ with $\NAND(c) = 0$.  We may assume $\alpha_v \neq 0$ since otherwise $\alpha_v/\alpha_p = 0$ and the inequality holds trivially.  Then, again dividing Eq.~\eqnref{e:hh} by $\alpha_v$,
using the induction hypothesis, and multiplying by $E$,
\begin{align}
-E h_{pv} \frac{\alpha_p}{\alpha_v} 
&= E \sum_c h_{vc} \frac{\alpha_c}{\alpha_v} - E^2 \nonumber \\
&\ge \sum_{c : \NAND(c) = 0} \frac{h_{vc}}{y_{0c}} - E^2 \Bigl(1 + \sum_{c : \NAND(c) = 1} h_{vc} y_{1c}\Bigr) \enspace .
 \label{e:Yyboundsfudge}
\end{align}
Let us upper-bound the coefficient of $E^2$:
\begin{align*}
1 + \sum_{c : \NAND(c) = 1} h_{vc} y_{1c}
&= 1 + \sum_{c : \NAND(c) = 1} h_{vc}^2 \sqrt{s_c} / \gamma_c \\
&\le 1 + \bigl( \max_c \frac 1 {\gamma_c} \bigr) \sqrt{s_v} \\
&\le (1 + \Gamma') \sqrt{s_v} \enspace ,
\end{align*}
where in the first inequality we used that $h_{vc} = (s_c / s_v)^{1/4}$ for all $c$ evaluating to $1$,
and that $\sum_c s_c = s_v$.

Now lower-bound the first term in Eq.~\eqnref{e:Yyboundsfudge}.  If $\NAND(c) = 0$, then
\begin{align*}
\frac { h_{vc} } { y_{0c} }
&= \frac { h_{vc}^2 } { 1 + \sum_d h_{cd} y_{1d} } \\
&\ge \frac { \sqrt{s_c/s_v} } { 1 + \sqrt{s_c} \max_d \frac 1 {\gamma_d} }
= \min_d \frac 1 {\sqrt{s_v}} \frac {\gamma_d} {1 + \frac {\gamma_d} {\sqrt{s_c}}} \\
&\ge \min_d \frac{\gamma_d}{\sqrt{s_v}} \left(1 - \frac{\gamma_d}{\sqrt{s_c}}\right)
\enspace ,
\end{align*}
where $d$ varies over the children of $c$.

Substitute these bounds back into Eq.~\eqnref{e:Yyboundsfudge}:
\begin{align*}
- E h_{pv} \frac { \alpha_p } { \alpha_v }
&\ge \frac 1 { \sqrt{s_v} } \left[ \min_{c,d} \gamma_d \left(1 - \frac{\gamma_d}{\sqrt{s_c}} \right) - E^2 (1 + \Gamma') s_v \right] \\
&\ge \frac 1 { \sqrt{s_v} } \min_{c,d} \left( \gamma_d - \frac{\Gamma^2}{\sqrt{s_c}} - E^2 (1 + \Gamma') s_v \right) \\
&= \frac 1 { \sqrt{s_v} } \min_{c,d} \left( \Gamma - \Gamma^2 \Big(\pathinvsum d + \frac1{\sqrt{s_c}}\Big)
    - E^2 (1 + \Gamma') \Big(\pathsum d + s_v\Big) \right) \enspace .
\end{align*}
This is at least $h_{pv} / y_{1v} = \gamma_v / \sqrt{s_v}$ for $\gamma_v$ defined as in Eq.~\eqnref{e:gammav}.
As $h_{pv}$, $y_{1v}$ and $E$ are all positive, we have $0 > \alpha_v/\alpha_p \geq -E y_{1v}$, as desired. \qedhere
\end{itemize}
\end{proof}

\begin{proof}[Proof of \thmref{t:spectralgap}]
Assume $\ket E$ is an eigenvector of $\H$ with energy $E \in (0,\frac{1}{9 \pathinvsum{r} \sqrt{\pathsum{r}}}]$.  We want to show $\alpha_{r'} = \alpha_{r''} = 0$.

Since $\NAND(r') = 1 - \NAND(r) = 1 - \varphi(x) = 0$, \lemref{t:ybounds} gives that either $\alpha_{r'} = \alpha_{r''} = 0$ or $0 < \frac{\alpha_{r''}}{\alpha_{r'}} \leq y_{0r'} E$.  In the latter case, Eq.~\eqnref{e:hh} at $r''$ gives $E \alpha_{r''} = h_{r''r'} \alpha_{r'}$, so
\begin{align*}
E^2
&\geq \frac { h_{r''r'} } { y_{0r'} } && \text{Subst. $y_{0r'}$ from Eq.~\eqnref{eqn:y1v}} \\
&= \frac{ h_{r''r'}^2 }{1 + h_{r'r} y_{1r}} && 
\begin{tabular}{@{}l}\text{Subst. $h_{r''r'} = 1 / (\sqrt{\pathinvsum{r}} N^{1/4})$, }\\ $\;$\text{$h_{r'r} = 1$, $y_{1r} = \sqrt{N} / \gamma_r$}\end{tabular} \\
&= \frac{ 1 / (\pathinvsum{r} \sqrt{N}) }{1 + \sqrt{N} / \gamma_r } && 1/\gamma_r \leq \Gamma' = 8 \pathinvsum{r} \\
&\ge \frac{1}{8 \pathinvsum{r}^2 N + \pathinvsum{r} \sqrt{N}}
\enspace .
\end{align*}
However, this is a contradiction, because $E^2 \le \frac 1 {81 \pathinvsum r ^2 N}$ due to $\pathsum{r} \ge N$ and $\pathinvsum{r} \ge 1$.
\end{proof}

\section{Coined quantum walks} \label{sec:szegedization}

In order to construct an algorithm from \thmref{t:eigensystem}, we need first to review briefly Szegedy's procedure for quantizing classical random walks.  \thmref{t:szegedization}, adapted from \cite{Szegedy04walkfocs}, relates the eigensystem of the coined quantum walk to that of the original classical walk.  

\def \tensor {\otimes}
\def \cH {{\cal{H}}}

\begin{theorem}[\cite{Szegedy04walkfocs}]
	\label{t:szegedization}
	Let $\{ \ket v : v \in V \}$ be an orthonormal basis for $\cH_V$.  For each $v \in V$, let $\ket{\tilde{v}} = \ket{v} \tensor \sum_{w \in V} \sqrt{p_{vw}} \ket{w} \in \cH_V \tensor \cH_V$, where $p_{vw} \geq 0$ and $\braket{\tilde v}{\tilde v} = \sum_w p_{vw} = 1$.  Let $T = \sum_v \ketbra{\tilde v}{v}$ and $\Pi = T T^\dagger = \sum_v \ketbra{\tilde v}{\tilde v}$ be the projection onto the span of the $\ket{\tilde v}$s.  Let $S = \sum_{v,w} \ketbra{v,w}{w,v}$, a swap.  Let $M = T^\dagger S T = \sum_{v,w} \ketbra{v}{\tilde v} S \ketbra{\tilde w}{w} = \sum_{v,w} \sqrt{p_{vw} p_{wv}} \ketbra{v}{w}$ a real symmetric matrix, and take $\{\ket{\lambda_a}\}$ a complete set of orthonormal eigenvectors of $M$ with respective eigenvalues $\lambda_a$.  
	
	Let $U = (2\Pi -1) S$, a swap followed by reflection about the span of the $\ket{\tilde v}s$.
	Then the spectral decomposition of $U$ is determined by that of $M$ as follows:
	Let $R_a = \Span\{T \ket{\lambda_a}, ST\ket{\lambda_a}\}$.  Then $R_a \perp R_{a'}$ for $a \neq a'$; let $R = \oplus_a R_a$.  
	$U$ fixes the spaces $R_a$ and is $-S$ on $R^\perp$.  The eigenvalues and eigenvectors of $U$ within $R_a$ are given by $\beta_{a,\pm} = -\lambda_a \pm i \sqrt{1 - \lambda_a^2}$ and $(1 + \beta_{a,\pm}S)T \ket{\lambda_a}$, respectively.
\end{theorem}

\begin{proof}
	First assume $a \neq a'$, and let us show $R_a \perp R_{a'}$.  Indeed, $\bra{\lambda_a} T^\dagger T \ket{\lambda_{a'}} = \braket{\lambda_a}{\lambda_{a'}} = 0$, as $T^\dagger T = 1$.  Since $S^2 = 1$, similarly, $S T \ket{\lambda_a}$ is orthogonal to $S T \ket{\lambda_{a'}}$.  Finally, $\bra{\lambda_a} T^\dagger S T \ket{\lambda_{a'}} = \bra{\lambda_a} M \ket{\lambda_{a'}} = 0$.  Therefore, the decomposition $\cH_V \tensor \cH_V = (\oplus_a R_a) \oplus R^\perp$ is well-defined.  
	
	$R$ is the span of the images of $S T$ and $T$.  $2\Pi-1$ is $+I$ on the image of $T$ and $-I$ on its complement; therefore $U$ is $-S$ on $R^\perp$.  
	
	Finally, $\Pi T = T T^\dagger T = T$ and $\Pi S T = T T^\dagger S T = T M$, so 
	\begin{align*}
		U (S T \ket{\lambda_a}) &= (2\Pi - 1) T \ket{\lambda_a} = T \ket{\lambda_a} \\
		U (T \ket{\lambda_a}) &= (2\Pi - 1)S T \ket{\lambda_a} = (2\lambda_a - S) T \ket{\lambda_a} \enspace ;
	\end{align*}
	$U$ fixes the subspaces $R_a$.  To determine its eigenvalues on $R_a$, let $\ket{\beta} = (1+ \beta S) T \ket{\lambda_a}$.  Then $U \ket{\beta} = (2 \lambda_a + \beta) T \ket{\lambda_a} - S T \ket{\lambda_a}$ is proportional to $\ket{\beta}$ if $\beta (2\lambda_a + \beta) = -1$; i.e., $\beta \in \{ -\lambda_a \pm i \sqrt{1-\lambda_a^2} \}$.  (If $\lambda_a = \pm 1$, note that $T \ket{\lambda_a} = \pm ST \ket{\lambda_a}$, so $R_a$ is one-dimensional with a single $U$ eigenvector.)
\end{proof}

To connect this theorem to classical and quantum walks, start with an undirected graph $G=(V,E)$.  Choose the $p_{v,w}$ to be the transition probabilities $v \rightarrow w$ of a classical random walk along this graph (i.e., with the constraint $p_{v,w} = 0$ if $(v,w) \notin E$).  Then $U = (2\Pi -1) S$ can be considered a quantization of the classical walk, taking place on the {\em directed edges} of $G$.  First the swap $S$ switches the direction of an edge.  Then, for the first register fixed to be $\ket{v}$, $(2\Pi-1)$ acts as a reflection about $\ket{\tilde v} = \ket{v} \tensor \sum_{w \sim v} \sqrt{p_{v,w}} \ket{w}$; it is a ``coin flip'' that mixes the directed edges leaving $v$.  
Therefore, although $U$ acts on $\cH_V \tensor \cH_V$, it preserves the subspace spanned by $\ket{v,w}$ and $\ket{w,v}$ for $(v,w) \in E$.  An alternative basis for this subspace is to give a vertex $v$ together with an edge index to describe an edge leaving $v$.  If the graph has bounded degree $\leq D$, then $U$ can be implemented on $\cH_V \tensor {\bf C^D}$ instead of $\cH_V \tensor \cH_V$.

Szegedy's \thmref{t:szegedization} relates the eigenvalues and
eigenvectors of the quantum walk $U$ to that of the matrix $M =
\sum_{v,w} \sqrt{p_{v,w} p_{w,v}} \ketbra{v}{w}$.  If $P =
\sum_{v,w} \sqrt{p_{v,w}} \ketbra{v}{w}$ is the element-wise
square-root of the transition matrix of the original classical
random walk, then $M$ is the element-wise product $P \circ P^T$.  $M$
can also be regarded as the Hamiltonian for a continuous-time quantum walk on the vertices of the underlying graph.

\subsection*{Discretization of general continuous-time quantum walks}

In our case, we are given $\H$ (from \secref{sec:hamiltonian}), and desire a factorization $\H = P \circ P^T$ such that $P$ has all row norms exactly one.  Then \thmref{t:szegedization} with $M = \H$ applies to relate the eigensystem of $\H$ to that of a certain coined quantum walk.  Such a factorization is possible for a large class of Hamiltonians: 

\begin{claim} \label{t:continuousdiscretecorrespondence}
Let $\H = \sum_{v,w} \H_{v,w} \ketbra{v}{w}$ be the positive-weighted symmetric adjacency matrix of a connected graph $G$.  Let $\ket \delta$ be the principal eigenvector of $\H$, with $\braket{v}{\delta} = \delta_v > 0$ for every $v$.  Assume $\norm \H = 1$.  Then $\H = P \circ P^T$, where $P = \sum_{v,w} \sqrt{\H_{v,w} \frac {\delta_w}{\delta_v}} \ket v \bra w$ has all row norms one.
\end{claim}

\begin{proof}
Since $\H$ is nonnegative, the principal eigenvector $\ket \delta$ is also nonnegative.  Since $\H$ is connected, $\delta_v > 0$ for every $v$; hence $P$ is well defined.  By construction, $P_{v,w} P_{w,v} = \H_{v,w}$, i.e., $P \circ P^T = \H$.  Furthermore, the squared norm of the $v$-th row of $P$ is $\sum_w P_{v,w}^2 = \frac 1 {\delta_v} \sum_w \H_{v,w} \delta_w = \frac {(\H \delta)_v} {\delta_v} = \norm \H = 1$, so $P$ corresponds to a classical random walk.
\end{proof}

\begin{remark} \label{t:continuousdiscretequantumwalkscorrespond}
Szegedy's \thmref{t:szegedization}, with \claimref{t:continuousdiscretecorrespondence}, serves as a general method for
relating an arbitrary continuous-time quantum walk on $G$'s vertices into a coined discrete-timed quantum walk on directed edges of $G$.  In
particular,
the eigenvalues of the discrete walk $-i U$ are given by $\pm \sqrt{1 -
\lambda_a^2} + i \lambda_a$ ($e^{i \arcsin \lambda_a}$ and $-e^{- i \arcsin \lambda_a}$)
whereas the continuous walk $e^{i M}$ has eigenvalues $e^{i \lambda_a}$.
The spectral gaps from zero of the continuous walk and the discrete walk are equal up to third order.
\end{remark}

\section{The algorithm} \label{sec:algorithm}

Recall from \defref{t:approxbalancedef} that a NAND formula is ``approximately balanced'' if $\pathinvsum{r} = O(1)$, $\pathsum{r} = O(N)$ and $\norm{\H} = O(1)$. 
These conditions
are satisfied, for example, by a balanced binary NAND tree, as $\pathinvsum{r} < 2$ and $\pathsum{r} < 2 N$ will then both be geometric series.  
The definition is also satisfied if for a fixed $\beta \in (0,\frac 1 2]$, for every vertex $v$ and every grandchild $c$ of $v$, $s_c \le (1-\beta) s_v$.  Under this condition, $\pathinvsum r = O(\frac 1 \beta)$ and $\pathsum r = O(\frac N \beta)$.  

\begin{theorem} \label{t:generaltheorem}
After efficient (i.e., $\poly(N)$ time) classical preprocessing of the size-$N$ formula $\varphi$ independent of $x$, $\varphi(x)$ can be evaluated with error $< 1/3$ using $N^{\frac{1}{2}+O(1/\sqrt{\log N})}$ queries to $U_x$.  The running time is also $N^{\frac{1}{2}+O(1/\sqrt{\log N})}$ assuming unit-cost coherent (oracle) access to the preprocessed string.  For a formula that is ``approximately balanced'' according to \defref{t:approxbalancedef}, the query complexity is only $O(\sqrt N)$ and the running time is only $\sqrt{N} (\log N)^{O(1)}$.
\end{theorem}

\begin{proof}

\subsubsection*{Preprocessing:}
If $\pathinvsum{r} \sqrt{\pathsum{r}} \norm{H} = N^{\frac12 + \Omega(1)}$, then 
preprocess the formula in two ways.  First, expand out gates so each NAND gate has $O(1)$ fan-in.  Since edge weights are all $\le 1$, this ensures that $\norm{H} = O(1)$.  Also, apply the formula ``rebalancing'' procedure of \cite{bce:size-depth, bb:size-depth} with parameter $k$ to be determined:

\begin{lemma}[{\cite[Theorem~4]{bb:size-depth}}] \label{t:rebalance}
    For all $k \geq 2$, one can efficiently construct an equivalent NAND formula $\varphi$
    with gate fan-ins at most two and
    satisfying\footnote{The constant in the depth bound is $9\ln 2$ instead of the $3 \ln 2$ in~\cite[Theorem~4]{bb:size-depth}, because we lose a constant converting an $\{ \textrm{AND, OR, NOT}\}$ formula to a NAND formula.}
    \begin{align*}
        \mathrm{depth}(\varphi) &\leq (9 \ln 2) k \log_2 N \\
        \mathrm{size}(\varphi) &\leq N^{1+1/\log_2 k} \enspace .
    \end{align*}
\end{lemma}

Let $\H$ be the Hamiltonian corresponding to a weighted adjacency matrix of the graph according to \defref{def:h}.  Compute a coined quantum walk operator $U$ that corresponds to $M = \H / n(\H)$ via \thmref{t:szegedization} (where $n(\H)$ is some upper bound on $\norm{\H}$, to ensure that all eigenvalues of $M$ have $\abs{\lambda_a} \leq 1$).
Obtaining $U$ takes a little care, since $\H$ depends on the oracle.  Consider $\H_{0^N}$ the Hamiltonian from \defref{def:h} assuming that all leaves evaluate to $0$.  By applying \claimref{t:continuousdiscretecorrespondence} as part of the preprocessing, we obtain a $U_{0^N}$ corresponding to $\H_{0^N} / \norm{\H_{0^N}}$.  Let $U = O_x U_{0^N}$, where $O_x$ is the input phase-flip oracle; conditioned on the current vertex being a leaf $i$, $O_x$ adds a phase of $(-1)^{x_i}$.  Then we claim that applying \thmref{t:szegedization} to $U$ will give $\H / \norm{\H_{0^N}}$.  Indeed, the only difference between $U$ and $U_{0^N}$ is on the $\ket{\tilde{i}}$s for leaf vertices $i$ with $x_i = 1$; in $U_{0^N}$,
$\ket{\tilde i} = \ket{i, p}$ 
($p$ being $i$'s parent), whereas in $U$, 
$\ket{\tilde i} = \ket{i, i}$ (i.e., $p_{i,i}=1$).
Therefore the $M$ from $U$ only differs from $\H_{0^N}$ in the coefficients involving leaves $i$ with $x_i = 1$, and $\bra{i} M \ket{p} = \bra{\tilde i} S \ket{\tilde p} = 0$; $M = \H / \norm{\H_{0^N}}$ as claimed.

\subsubsection*{Algorithm:}

\begin{enumerate}
\item
Prepare $\ket{\tilde r''} = \ket{r'', r'}$.
\item
``Measure the energy under $\H$."  That is, apply quantum phase estimation to $-i U = -i O_x U_{0^N}$.  Use precision $\delta_p ={1}/ ({10 \pathinvsum{r} \sqrt{\pathsum{r}}})$ and error probability $\delta_e$ any constant less than $1/4$.

\item
Output zero if and only if the measured phase is $0$ or $\pi$.
\end{enumerate}

\figref{f:algorithm} lays out the steps of the algorithm in complete detail for the case of a balanced binary NAND tree.  We did not use \claimref{t:continuousdiscretecorrespondence} to derive $U$ in \figref{f:algorithm}, because in this special case it is clear that applying \thmref{t:szegedization} to $U$ gives $\H$, except with larger weights to leaves evaluating to $0$ (see \remref{t:alternateweights}).

\subsubsection*{Correctness:}

The correctness follows from Theorems~\ref{t:eigensystem} and~\ref{t:szegedization}.  If $\varphi(x) = 0$, then there exist two eigenvectors of $U$ given by $(1 \pm i S) T \ket{a}$, with eigenvalues $\pm i$, respectively.  Their overlaps with the initial state $\ket{\tilde r''}$ are $\abs{\bra{\tilde r''} (1 \pm i S)T \ket{a}} = \abs{a_{r''} \pm i a_{r'} \frac{ h_{r''r'} }{\norm{H} }} \geq 1/\sqrt 2 - O(1/N^{1/4})$.  Since the norm of $(1 \pm i S)T\ket{a}$ is at most $2 \norm{\ket{a}} = 2$, we find that the probability of outputting $0$ is at least $2 \big( \tfrac12 (\frac1{\sqrt{2}}-o(1))\big)^2 = 1/4 - o(1)$.

Conversely, if $\varphi(x) = 1$, then every eigenstate of $\H$ with
support on $r'$ or $r''$ has energy at least $1 / (9
\pathinvsum{r} \sqrt{\pathsum{r}})$ in magnitude.  Every eigenstate
of $U$ with support on $\ket{r'',r'} = \ket{\tilde r''} = T
\ket{r''}$ must be of the form $(1 + \beta_{a,\pm}S)T
\ket{\lambda_a} = (1 + (-\lambda_a \pm i \sqrt{1-\lambda_a^2}) S)T
\ket{\lambda_a}$.  The terms which can overlap $T \ket{r''}$ are
either $\braket{r''}{\lambda_a}$ (via $T$) or
$\braket{r'}{\lambda_a}$ (via $S T$).  But for
$\abs{\lambda_a} < 1 / (10 \pathinvsum{r} \sqrt{\pathsum{r}})$, 
both coefficients must be zero.  Therefore, our algorithm outputs
$0$ with probability less than $\delta_e < 1/4$.  This constant gap
can be amplified as usual.  

\subsubsection*{Query and time complexity:}

Phase estimation requires applying $O(\norm{\H_{0^N}} / (\delta_p \delta_e)) = O(\pathinvsum{r} \sqrt{\pathsum{r}} \norm{\H_{0^N}})$ controlled-$U$ evolutions~\cite{cemm:qalg}.  For an approximately balanced graph, this is $O(\sqrt{N})$.  For a general graph, the number of controlled-$U$ applications is $O(\sqrt{s_r} d_r^{3/2} \norm{H_{0^N}})$ due to the bounds on $\sigma_\pm(r)$ from \defref{t:pathsum}.  For a rebalanced formula from \lemref{t:rebalance} with parameter $k$, this is $O(N^{\tfrac12+\tfrac1{2\log_2 k}} (k \log_2 N)^{3/2})$ since $\norm{H} = O(1)$.
Set $k = 2^{\sqrt{\frac {\log_2 N} 3}}$ to optimize this bound to be $O(N^{\tfrac12 + \sqrt{(3+o(1))/\log N}})$ queries to $O_x$.

During the preprocessing phase, for each vertex $v$ we compute a sequence of $O((\log N)^2 \log \log N)$ elementary gates that approximate to within $1/N$ the coin diffusion operator at $v$, by applying the Solovay-Kitaev Theorem~\cite{ksw:qc-book}.  (Using these approximations, the algorithm's total error probability will only increase by $o(1)$.)  Store the descriptions of these gate sequences in a classical string, which we assume the algorithm can access coherently at unit cost.  The algorithm at vertex $v$ looks up the corresponding gate sequence and applies it to the coin register $\ket c$.  The total running time is thus only polylogarithmically larger than the number of queries.
\end{proof}

\section{Open problems}

We conclude by mentioning some open problems.
\begin{itemize}
\item
Our algorithm needs to know the full structure of the formula beforehand to determine the coin's bias at each internal vertex.  (Coin biases are determined by the principal eigenvector of the graph's weighted adjacency matrix; they can also be solved for recursively from the leaves to $r$.)  However, these coefficients need not be computed exactly, because there is some freedom in the recurrence on $y_{0v}$ and $y_{1v}$.  It would be interesting to know if a different choice of coefficients, or a relaxed calculation thereof, would allow for faster preprocessing.  Furthermore, it would be interesting to know on what kinds of structured inputs the preprocessing can be done in time $N^{\frac 1 2 + o(1)}$.
\item
Some numerical simulations indicate that the formula can be evaluated by running the quantum walk from the initial state, and measuring whether the quantum state has large overlap with $\frac 1 {\sqrt 2} (\ket{r',r''} + \ket{r'',r'})$ or $\frac 1 {\sqrt 2} (\ket{r',r''} - \ket{r'',r'})$.
If this is true, then we can avoid the phase estimation on top of the coined quantum walk, simplifying the algorithm.
\item
What kinds of noisy oracles can this algorithm, or an extended version, tolerate?
For example, \cite{hmw:berror-search} extends Grover search to the case where input values are computed by a bounded-error quantum subroutine.  
\item
Are there hard instances of formulas for which the rebalancing provided by \lemref{t:rebalance} is tight?  Are these also hard instances for our algorithm?
For example, the most imbalanced formula, $\varphi(x_1 \dots x_N) = x_1 \NAND (x_2 \NAND (x_3 \NAND (\dots \NAND x_N)))$, is not a hard instance.  It can be rebalanced by a different procedure into depth $O(\log N)$ and size $O(N \log N)$, and can be evaluated with $O(\sqrt N)$ queries.
\end{itemize}

\section*{Acknowledgments}

Robert would like to thank the Institute for Quantum Information at Caltech for hospitality.
We thank Ronald de Wolf for comments on an early draft of the paper, and thank J\'er\'emie Roland for thoughtful remarks.

\bibliographystyle{alpha}
\bibliography{andor}

\end{document}